\documentclass[preprint2]{aastex}
\def\eg{\hbox{e.g.}}

\def\etal{\hbox{et~al.}}

\def\vinf{\hbox{$v_\infty$}}
\def\Mdot{\hbox{$\dot M$}}
\def\Rstar{\hbox{R$_*$}}
\def\Rrat{\hbox{$\cal R$}}
\def\Grat{\hbox{$\cal G$}}
\def\zori{\hbox{$\zeta$ Ori}}
\def\kms{\hbox{km$\,$s$^{-1}$}}
\def\ne{\hbox{$n_e$}}

\def\Teff{\hbox{$T_{eff}$}}
\def\NW{\hbox{$N_w$}}
\def\cmm3{\hbox{cm$^{-3}$}}

\def\CIV{\hbox{C\,{\sc iv}}}
\def\OVII{\hbox{O\,{\sc vii}}}
\def\OVIII{\hbox{O\,{\sc viii}}}
\def\NEIX{\hbox{Ne\, {\sc ix}}}
\def\NEX{\hbox{Ne\, {\sc x}}}
\def\NEIX{\hbox{Ne\, {\sc ix}}}
\def\MGXI{\hbox{Mg\,{\sc xi}}}
\def\MGX{\hbox{Mg\,{\sc x}}}

\def\SIXIII{\hbox{Si\,{\sc xiii}}}
\def\SXV{\hbox{S\,{\sc xv}}}

\def\Dlamd{\hbox{$\Delta \lambda_D$}}

\shorttitle{High Density X-ray Plasma on \zori}
\shortauthors{Waldron \& Cassinelli}

\slugcomment{To be published in {/it The Astrophysical Journal
Letters} (Jan or Feb 2001)}

\begin{document}

\title{Chandra Discovers a Very High Density X-ray Plasma on the O-Star $\zeta $ Orionis}

\author{Wayne L. Waldron\altaffilmark{1} \& Joseph P.
Cassinelli\altaffilmark{2}}

\altaffiltext{1}{Emergent Information Technologies, Inc., Space Sciences
Division, 9315 Largo Drive West, Suite 250, Largo, MD 20774;
wayne.waldron@emergent-IT.com}
\altaffiltext{2}{Astronomy Department, University of Wisconsin,
475 N. Charter St., Madison, WI  53706; cassinelli@astro.wisc.edu}

\begin{abstract}

We report on a Chandra line spectrum observation of the O supergiant,
$\zeta$ Orionis (O9.7 Ib).  A 73.4 ks HETGS observation shows a wide range
of ionization stages and line strengths over the wavelength range of 5 to 26
\AA. The observed emission lines indicate a range in temperature of 2
to 10 MK which is consistent with earlier X-ray observations of \zori. Many
lines are spectrally resolved showing Doppler broadening of $900 \pm 200$
\kms. The observed He-like ions (O VII, Ne IX, Mg XI, and Si XIII) provide
information about the spatial distribution of the X-ray emission.  Although
the observations support a wind distribution of X-ray sources, we find three
conflicting results.  First, line diagnostics for \SIXIII\ indicate that
this line emission forms very close to the stellar surface, where the
density is of order $10^{12}$ cm$^{-3}$, but the velocity there is too small
to produce the shock jump required for the observed ionization level.
Second, the strong X-ray line profiles are symmetric and do not show any
evidence of Doppler blue-shifted line centroids which are expected to
accompany an outwardly moving source in a high density wind.  Third, the
observed velocity dispersions do not
appear to correlate with the associated X-ray source radii velocities,
contrary to expectations of wind
distributed source models. A composite source model involving wind shocks
and some magnetic confinement of turbulent hot plasma in a highly
non-symmetric wind, appears to be needed to explain the line diagnostic
anomalies.

\end{abstract}

\keywords{X-rays: stars -- Line: profiles-- stars: early-type--
stars: individual (\zori)-- stars: magnetic fields-- stars: mass-loss}

\section{Introduction}

The Orion Belt supergiant \zori\ (O9.7 Ib) is one of the most thoroughly
studied hot stars.  It has been previously observed in X-rays with the
Einstein, ROSAT, and ASCA satellites. Observations with the Einstein Solid State
Spectrometer (SSS) found it to be the first O-star to show line
emission at \SIXIII\ and \SXV\ (Cassinelli \& Swank 1983).  ASCA
observations show that \zori\ may have a weak high temperature component
greater than 10 MK (Waldron, Corcoran, \& Kitamoto 1997), but the majority
of the X-ray emission is soft, peaking below 1 keV.  The observed X-ray
emission is consistent with the standard O-star relation: L${_X} \sim
10^{-7}$ L$_{Bol}$ (Bergh\"{o}fer et al. 1997), and has remained
relatively stable except for a short term of $\sim 15 \%$ variability
observed during a ROSAT monitoring study by Bergh\"{o}fer \& Schmitt (1994).

It is generally believed that the hot star X-ray emission arises from
shocks that develop in the outflowing stellar wind owing to instabilities
associated with the line driven wind mechanism (Lucy \& White 1980; Owocki,
Castor, \& Rybicki 1988; Feldmeier 1995).  The original X-ray models for hot
stars proposed that the X-rays arise from a thin coronal structure located
at the base of the stellar wind (Cassinelli \& Olson 1979; Waldron 1984).
However, the base coronal model was considered inappropriate because early
observations did not show significant attenuation of soft X-rays at the
oxygen K-shell edge which was expected due to the overlying high density
cool wind (Cassinelli \& Swank 1983).

The X-ray emission lines produced by a stellar wind distribution of
outwardly propagating shocks should show broad profiles with a shape that
depends on the wind optical depth (MacFarlane \etal\ 1991). For an optically
thin wind, the predicted profiles are flat-topped and symmetric, whereas,
for an optically thick wind, the profile becomes asymmetric, showing a
blue-shifted, roughly triangular-shape, owing to the larger attenuation of
the line X-rays from the far side of the stellar envelope. The net degree of
asymmetry is determined by the number of shocks and the column density, \NW,
of the overlying stellar wind. The blue-ward extent of the line reflects the
largest outward velocities of the shock fronts.

The Chandra spectrometers have the necessary spectral resolution and sensitivity to test these
shock model predictions. In this Letter we present a preliminary analysis of the spectral lines from
our Chandra High-Energy Transmission Grating Spectrometer (HETGS) observation of \zori.

\section{Observations and Analysis}

Our Chandra HETGS observation of \zori\ was performed in two parts: on 2000
April 8 (04$^{h}$08$^{m}$ - 21$^{h}$37$^{m}$ UT), and on 2000 April 9
(20$^{h}$38$^{m}$ - 01$^{h}$05$^{m}$ UT), with effective exposure times of
59.7 ks and 13.7 ks respectively. \zori\ is the primary star of a widely
separated binary system with an early B-star giant companion, and with
Chandra the two components are cleanly separated. Here we use the standard
pipeline processed data from our combined 73.4 ks observation to discuss the
X-ray line spectrum of \zori\ A. The high-energy grating (HEG) and the
medium-energy grating (MEG) spectra are shown in Figure 1.  The range of
ions identified indicate an X-ray temperature range of $\sim 2$ to 10 MK.
The triads of He-like lines, commonly known as the ``$fir$ lines'' (see Sec
2.1), of
\OVII, \NEIX\, \MGXI\, and \SIXIII\ are identified in Figure 1.

In this Letter, we focus on the two most obvious features observed in the HETGS
spectrum of \zori; (1) the unusual line emission properties of He-like ions,
and; (2) the unexpected combination of large line widths and negligible line
centroid shifts.

\scalebox{0.4}[0.4]{\rotatebox{180}{\includegraphics{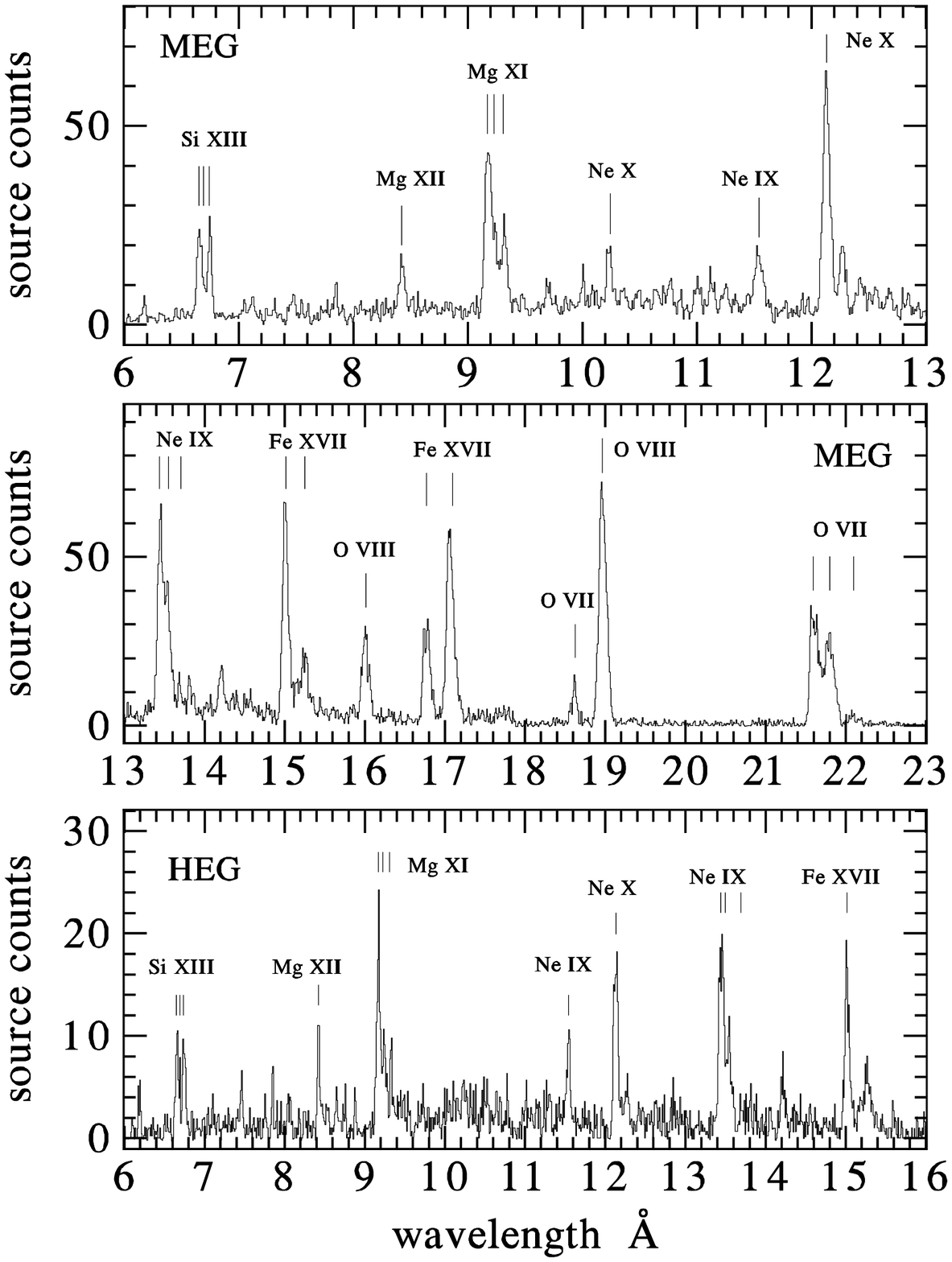}}}
\figcaption[]{Co-added $1^{st}$ order, background subtracted, Chandra HETGS
MEG (top panels) and HEG (bottom panel) count spectra of \zori.  The ions
responsible for the strongest line emissions are identified. The bin size is
0.01 \AA.
\label{fig1}}

\subsection{Analysis of He-like Spectral Line Transitions}

Gabriel \& Jordan (1969) were the first to demonstrate that the $fir$
(forbidden, intercombination, and resonance) emission lines of He-like ions
can provide very useful diagnostics for the study of X-ray emitting plasmas.
They found that the line flux ratio $\Rrat = f/i$ is density sensitive, and
the ratio $\Grat = (f+i)/r$ is temperature sensitive.  Although these $fir$
line diagnostics have been used extensively in solar studies to derive X-ray
electron densities ($n_e$) and temperatures ($T_e$) (\eg, Wolfson et al.
1983), the application to O-star spectra requires special care.  Blumenthal,
Drake, \& Tucker (1972), showed that the presence of an external strong UV
radiation field can lead to overestimates of the density, since not only
collisions, but radiative excitations as well can lead to a decrease in the
strength of the $f$ line relative to the $i$ line.  Hence, for O-stars, the
observed \Rrat\ values for several He-like ions, may no longer lead to a
direct measure of $n_e$, but instead provides information on the radial
distance of the X-ray source from the star.

Using the formalism developed by Blumenthal et al. (1972), we calculate the
radial dependence of \Rrat\ in the envelope of \zori\ using a photospheric
UV flux model at \Teff = 30,000K (Kurucz 1993). The radial dependent wind density
is derived from the stellar and wind parameters of Lamers \& Leitherer
(1993) using a standard wind velocity law that is modified below 1.02
stellar radii to provide a smooth density transition to the photospheric
structure. The radial dependence of the UV mean intensity $J_{\nu}(r)$ is
assumed to be geometric dilution of the stellar radiation, with no
attenuation, i.e., $J_{\nu}(r) = 4~W(r) H_{\nu}(\Teff) $, where $ W(r) =
\frac{1}{2} \left( 1 - \sqrt{1-(\Rstar/r)^2} \right)$, and $H_{\nu}$ is given by Kurucz (1993).
The predicted \Rrat\ dependencies for \OVII, \NEIX, \MGXI, and \SIXIII\ are shown in Figure 2.

\scalebox{0.45}[0.45]{\rotatebox{0}{\includegraphics{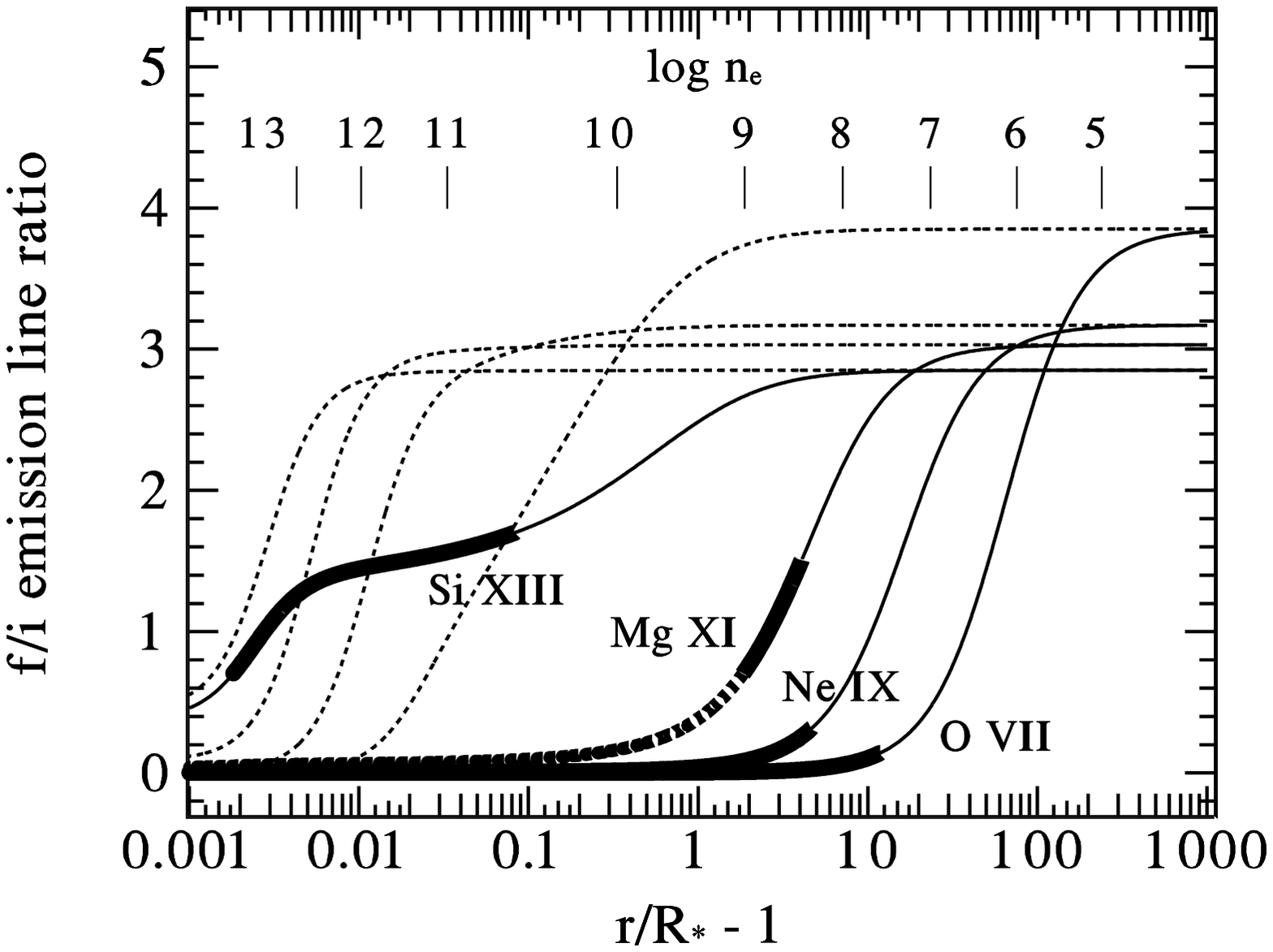}}}
\figcaption[]{The \OVII, \NEIX, \MGXI, and \SIXIII\ $f/i$ ratio dependence on
geometric dilution (radius) of the UV radiation field and electron density
(vertical lines) for the circumstellar environment of \zori.  The solid
lines include the effects of a UV radiation field (appropriate for hot
stars), whereas the dashed lines neglect these effects.  The range in the
observed $f/i$ ratios are indicated by heavy solid lines.  The heavy dashed
line for \MGXI\ is an extension owing to possible overestimates of the $f$
line strength because of contamination from a \MGX\ line. The corresponding
wind densities associated with the radius scale are shown at the top of the
panel.
\label{fig2}}

\newpage

For comparison, we also show the \Rrat\ dependence for the collision only case, appropriate for
very low UV flux.

We find that the predicted \Rrat\ ratios for the Chandra observed He-like
ions, except \SIXIII, are controlled by the strength of the UV radiation
field, independent of density, as indicated in Figure 2 by the large
differences between the collisional and radiative curves.  However, for
\SIXIII\ the predicted \Rrat\ ratio shows three regions of dominance,
collisional ($n_e > 10^{13}$ cm$^{-3}$), constant geometric dilution $( <
1.1 \Rstar) $, and a decreasing geometric dilution for large radii.  This
different \Rrat\ dependence for \SIXIII, as compared to the other three
He-like ions, is related to the different strengths of the UV radiation
field on the two sides of the Lyman jump. In the case of \SIXIII\ the
photo-excitation between the upper levels of the $f$ and $i$ lines ($2^3S
\rightarrow 2^3P$) is produced by radiation shortward of 912 \AA, where the
emergent stellar flux is $\sim$ 8 times smaller than that longward of 912
\AA. Figure 2 shows that \SIXIII\ marginally allows for a direct estimate
of \ne. Since the other He-like ions have their excitation transition
longward of 912 \AA, the observed \Rrat\ ratio can only lead to a measure
of the geometrical dilution factor (or radius).

Figure 2 shows the measured range of \Rrat\ values (given as the heavy solid
lines) as a function of radius. The \Rrat\ values for \OVII\ and \NEIX\ are
extracted from the MEG spectrum, whereas, for \MGXI\ and \SIXIII, the HEG
spectrum is used. From Figure 2 we can read off the radii of the X-ray
emission line regions, and the associated model dependent electron densities. The
most interesting result is the observed \Rrat\ ratio for \SIXIII\ of $1.2
\pm 0.5$ which indicates that this line emission originates from the base
of the wind, where the electron density is of order $10^{12}$ to $10^{13}$
cm$^{-3}$. The \MGXI\ \Rrat\ ratio gives a range in the formation radius of
3 to 5 \Rstar, and $n_e$ of order $10^9$ cm$^{-3}$. The heavy dashed line for
\MGXI\ in Figure 2 represents the possibility that the $f$ line may be
contaminated by a \MGX\ line since the HEG \MGXI\ $f$ line shows evidence of
a red-shift of $\sim$ 0.015 \AA.  The other two ions,
\OVII\ and \NEIX, have very weak $f$ emission lines, hence, their \Rrat\ ratios can only provide
upper limits on the source radii of 12 and 6 \Rstar, corresponding to lower limits on $n_e$ of $ 4
\times 10^{7}$ cm$^{-3}$, and $ 2 \times 10^{8}$ cm$^{-3}$ respectively.
The X-ray location from \MGXI\ and upper limits from \OVII\ and \NEIX\ shown
in Figure 2 suggest that some of the X-rays are produced high in the wind,
consistent with stellar wind shock predictions (Feldmeier, Puls, \&
Pauldrach 1997). Furthermore, shocks may still be generated above these
upper limits, but they cannot be the dominant source of the He-like X-ray
emission lines, otherwise their $f$ line fluxes and \Rrat\ ratios would
be larger than observed.

The X-ray source radii predicted from the $f/i$ analysis (Fig. 2) suggest a
systematic ordering of the He-like ions, i.e., high energy ions at low radii
and low energy ions at large radii.  In a shock picture, a reverse ordering
would have been expected, since the shock jump temperature should be proportional to the local
wind velocity.  This particular anomaly might be explainable by
considering the radial distance from which the line emission can be seen by
an external observer. Using the wind absorption cross sections of Waldron et
al. (1998), we find that the optical depth in the continuum adjacent to a
line reaches unity at radii of 1.03, 1.5, 2.8, and 2.2 for \SIXIII, \MGXI,
\NEIX, and \OVII, respectively. These radii are all just less than or equal to
the radii predicted by our $f/i$ ratio analysis.  We suggest, therefore,
that our Chandra observations imply that the X-rays emission lines originate
primarily from just above their associated monochromatic X-ray "photosphere"
where the continuum radial optical depth is unity.

Finally, using the results of Pradhan \& Shull (1981), Wolfson et al.
(1983), and Keenan et al. (1994), we extract $T_e$ from the measured \Grat\
ratios.  For \OVII\ and \NEIX, we find upper limits of 2 and 6 MK, and for
\MGXI\ and \SIXIII, ranges of 2 - 11 and 1 - 9 MK are indicated. These
temperatures are consistent with the overall observed ion distribution
illustrated in Figure 1.

\subsection{Analysis of the Emission Line Profiles}

After correcting for instrumental broadening, all X-ray lines in \zori\ are
seen to be very broad, with velocity dispersions in the range $900 \pm 200$
\kms. Our HETGS observations of \zori\ show no evidence of any
large Doppler blue-shifted asymmetric X-ray line profiles.
We also find no indication that the Gaussian width correlates with the ion
properties,  which suggests that the
lines form in the same region or at least in regions with similar velocity
spreads.  From the strongest HEG lines, we find a distribution in shifts
from line center of $\pm 0.005$ \AA\ which is consistent with the wavelength
uncertainty of the HETGS/HEG (Marshall 2000).  There are a few MEG spectral
lines showing blue-shifts of $\sim$ 0.01 \AA\ (e.g., \OVIII, 18.97 \AA),
but we also see similar red-shifts.  In this preliminary analysis we find
that there is no observed preference for a systematic blue or red-shift.

Our $f/i$ analysis in Sec. 2.1 indicates a distribution of X-ray sources
from $\sim$ 1 to 12 \Rstar. However, our observed line profiles appear to show very similar
line-width and line-symmetry properties, which seems
inconsistent with their being formed over such a large range in radius. To
better understand this problem, we explore two distributed shock models. We
only focus on those X-ray lines that form above $\sim$ 1.5 \Rstar\ which
arise from shocks in the wind (i.e., \MGXI, \NEIX, and \OVII).

Expanding on the single shock model of MacFarlane et al. (1991), we
developed a shock model consisting of 10 spherically symmetric shocks
equally distributed between 0.4 to 0.97 \vinf, with temperatures ranging
from 2 to 10 MK.  The shock strength scales with the local wind emission
measure, and the wind absorption cross sections are from Waldron et al.
(1998). This choice of parameters generates an X-ray source distribution
with radius that is consistent with our $f/i$ analysis. The model
predictions are shown in Figure 3a for the two limiting cases, optically thick (maximum \NW),
and optically thin ($\NW = 0$).  In the optically thick
case (appropriate for a high density hot star wind), the wind
attenuation reduces the contribution to the line emission from the far side
of the star and the resultant line emission tends to arise from the highest density regions that
are blue-shifted and moving towards us.  For the optically thin case, the
line emission is narrow and almost symmetric (with a slight asymmetry due to
stellar occultation effects).  The narrowness arises because emission is
proportional to $ n_e^2 $, thus the strongest line emission occurs from
regions of low velocity. Comparison of the shock model predictions (Fig. 3a),
folded through the MEG instrumental response, with a typical single
\zori\ X-ray line profile (\NEX, 12.13 \AA) is shown in Figure 3b, along
with the best-fit Gaussian model (\Dlamd = $0.042 \pm 0.003$ \AA).  Clearly
the optical thick line model does not explain the observed line profile, and
the optically thin case provides a good fit to the observed symmetry
and broadness. However, the optically thin model requires extreme conditions.

\scalebox{0.4}[0.4]{\rotatebox{180}{\includegraphics{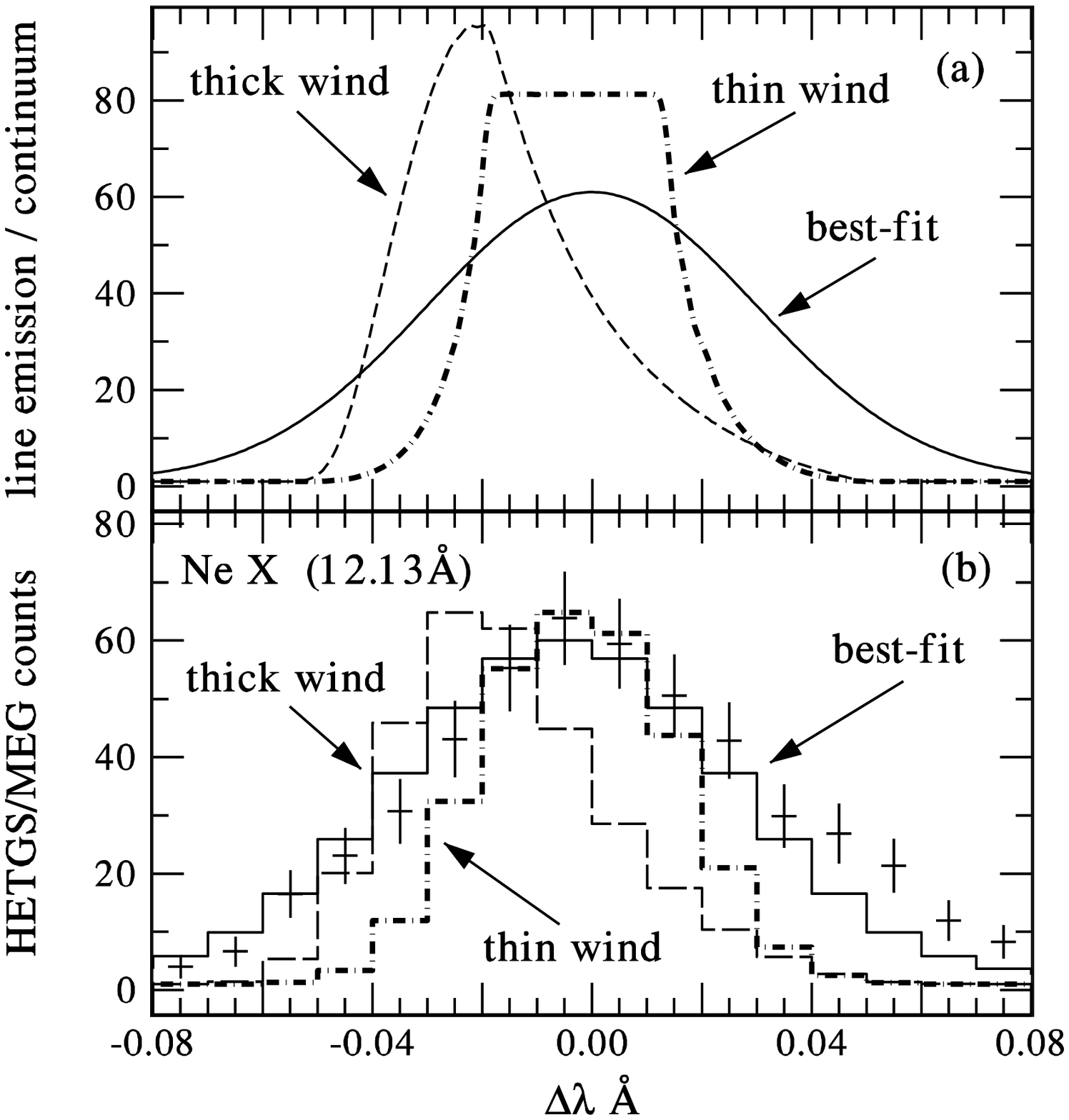}}}
\vspace{-1.5cm}
\figcaption[]{X-ray line profile dependence on wavelength shift
({$\Delta \lambda$}) from line center showing 3 input line profiles and
their expected HETGS/MEG count spectra. (a) Model line profiles for a
distribution of shocks for an optically thick (dashed line) and thin wind
(dot-dashed line).  The best-fit Gaussian model profile (solid line) is
shown for comparison which required a physical velocity dispersion of $1040 \pm 75$
\kms. (b) Comparison of the observed Ne X line at 12.13 \AA\ (1 sigma error
bars) with model line profiles folded through the instrumental response of
the HETGS/MEG.
\label{fig3}}
\vspace{1cm}
The wind could be made thin by increasing the ionization structure, which
reduces the opacity (Waldron 1984; MacFarlane, Cohen, \& Wang 1994). However,
for normal hot stars, the winds cannot be uniformly highly ionized because
these stars show strong P-Cygni UV line profiles from low ion stages such as
\CIV, which are distributed throughout the wind (Lamers \& Morton 1976).  An
optically thin wind can also be produced by reducing the mass loss rate,
\Mdot\, since \NW\ is proportional to \Mdot. For example, to maintain an
X-ray line symmetry comparable to the optically thin case shown in Figure
3a, while constraining the sources to lie within 12 \Rstar, \Mdot\ must be
reduced by at least an order of magnitude. However, such a reduction in
\Mdot\ would be inconsistent with wind line profile analyses ( Lamers \& Morton
1976) and free-free continuum observations (Lamers, Waters, \& Wesselius
1984, Abbott et al. 1980). There are other alternatives.  For example, a
wind with a clumpy or non-symmetric structure, has been shown to effectively
produce an optically thin wind even in stars with relatively large \Mdot\
(Moffatt \& Robert 1994).  In addition, if the shocks are produced at very
large stellar radii where the corresponding column densities between equal
red and blue-shifted velocities components are negligible, a symmetric line
profile can be produced (Feldmeier, private communication).  However, to
achieve a difference of no more than 10\% between the red and blue-shifted
velocity components for the \NEX\ line shown in Figure 3b, we find that the
shock would have to be located at $\sim$ 25 \Rstar\ which is $\sim$ 2 times
larger than the largest formation radius inferred from our $f/i$ analysis.

\section{Discussion}

Our HETGS observation of \zori\ has presented several unexpected results.
Although the analysis of the He-like ions suggest a stellar wind
distribution of X-rays, consistent with a shock model, we also find
convincing evidence for X-ray emission from near the stellar surface. Such a
source location presents a problem because the wind velocity would be too
small to produce a shock jump capable of producing the \SIXIII\ ion stage.
The observed X-ray line profiles are very broad and symmetric, and they show
no evidence for Doppler blue-shifted line centroids.  Since the observations
indicate that the shocks must be located within 12 \Rstar, the observed line
profiles imply that a significant reduction in \NW\ or \Mdot\ is required to
explain the observed line symmetry. Even if we can explain the line symmetry
problem by applying non-symmetric or clumpy wind structure arguments, we
still need an alternative model to explain the origin of the high density,
near-surface X-ray plasma from \zori.  We suggest that there is magnetic
loop confinement as was originally proposed by Cassinelli \& Swank (1983) to
explain their SSS observations of \zori.

The observed \zori\ surface X-ray plasma density is about 2 orders of
magnitude larger than normal for a quiescent solar magnetic loop (e.g.,
Vaiana \& Rosner 1978), but is comparable to solar flare densities (Phillips
et al. 1996).  A surface distribution of magnetic loop structures could
provide a mechanism for producing a collection of symmetric blue and
red-shifted bulk plasma motion, and since both components are near the
surface of the star, their emitted X-rays must pass through similar amounts
of absorbing material, producing symmetric line profiles like those observed
in \zori.  Although the observed velocity dispersions are high, they are
comparable to predicted solar flare velocity dispersions of order 1000 \kms\
(Antonucci, Benna, \& Somov 1996). If the high density plasma of \zori\ is residing in a collection
of surface magnetic loops, we can estimate the required magnetic field by assuming a
balance between the magnetic and gas pressures (i.e., the plasma $\beta$ is
unity).  Using an $n_e$ of $10^{12}$ cm$^{-3}$ and a $T_e$ of 5 MK leads to
a magnetic field strength $> 180$ Gauss.  This is comparable to the field of
$\sim$ 200 Gauss detected on the B1 giant $\beta$ Cephei (Henrichs et al.
2000). The likelihood of such fields on hot stars has been investigated by
Cassinelli \& MacGregor (1999) who found that fields in hot main sequence
stars generated by dynamo action at the interface between the radiative core
and convective envelope could rise to the surface. Thus a magnetically
confined plasma on the surface of a hot star is certainly plausible.

Because of the $n_e^2$ dependence of recombination that leads to the He-like
ions, the $fir$ diagnostic of density or of source location tend to yield
values for the densest region producing the lines.  This was discussed for
the density dominated case by Brown \etal\ (1991). We find evidence that
X-rays are being produced both at dense regions very near the star, and from
regions distributed throughout the wind.  However, the observed X-ray {\it line profiles} seem
incompatible with their being distributed as shocks
throughout the optically thick wind. A more fundamental question
that we raise is - why are the velocity dispersions essentially the same
regardless of the X-ray source location? Our results provide both new
facts and new challenges for future studies of hot star X-ray emission.

\acknowledgements

We are grateful to Mike Corcoran, Andy Endal, Joe MacFarlane, and Nathan
Miller for their useful comments and suggestions.  We would also like to
thank John Brown for several enlightening conversations, and the referee,
Achim Feldmeier, for his detailed critique of our manuscript, and several
informative suggestions.

\end{document}